\documentclass[pra,twocolumn,showpacs]{revtex4}
\usepackage{amsmath,amssymb,mathrsfs}
\usepackage{psfrag}
\usepackage{graphicx}
\usepackage{graphics}
\usepackage{epsfig}
\usepackage{bm}
\usepackage{color}
\usepackage{verbatim,color,ulem}

\newcommand{\beq}{\begin{equation}}
\newcommand{\eeq}{\end{equation}}
\newcommand{\beqa}{\begin{eqnarray}}
\newcommand{\eeqa}{\end{eqnarray}}

\begin{document}

\title{Thermal field theory of bosonic gases 
with finite-range effective interaction}
\author{A. Cappellaro$^{1}$ and L. Salasnich$^{1,2,3}$}
\affiliation{$^{1}$Dipartimento di Fisica e Astronomia ``Galileo Galilei'', 
Universit\`a di Padova, Via Marzolo 8, 35131 Padova, Italy
\\
$^{2}$Consorzio Nazionale Interuniversitario per le Scienze Fisiche 
della Materia (CNISM), Unit\`a di Padova, Via Marzolo 8, 35131 Padova, Italy
\\
$^{3}$Istituto Nazionale di Ottica (INO) del Consiglio Nazionale 
delle Ricerche (CNR), Via Nello Carrara 1, 50019 Sesto Fiorentino, Italy}

\date{\today}

\begin{abstract}
We study a dilute and ultracold Bose gas of interacting atoms by using 
an effective field theory which takes account 
finite-range effects of the inter-atomic potential. Within the formalism of 
functional integration from the grand canonical partition function 
we derive beyond-mean-field analytical results which depend 
on both scattering length and effective range of the interaction. 
In particular, we calculate the equation of state of the bosonic system 
as a function of these interaction parameters 
both at zero and finite temperature including one-loop Gaussian fluctuation. 
In the case of zero-range effective interaction we explicitly show 
that, due to quantum fluctuations, the bosonic system is thermodynamically 
stable only for very small values of the gas parameter. 
We find that a positive effective 
range above a critical threshold is necessary to remove the 
thermodynamical instability of the uniform configuration. 
Remarkably, also for relatively large values of the gas parameter, 
our finite-range results are in quite good 
agreement with recent zero-temperature Monte Carlo calculations obtained 
with hard-sphere bosons. 
\end{abstract}

\pacs{03.75.Ss 03.70.+k 05.70.Fh 03.65.Yz} 

\maketitle

\section{Introduction} 

The experimental achievement of Bose-Einstein condensation with dilute 
and ultracold alkali-metal atoms \cite{anderson1995,bradley1995,davis1995} 
has triggered many theoretical investigations of 
weakly-interacting Bose gases. 
The key theoretical tool for the description of these bosonic systems  
at zero temperature is the Gross-Pitaevskii equation \cite{gross}, 
where the nonlocal inter-atomic interaction is approximated by a local 
contact interaction characterized by only one physical parameter, the s-wave 
scattering length $a_s$. Also quantum and thermal fluctuations 
have been analyzed within this contact approximation of the 
inter-atomic potential (see for instance 
\cite{shi1998,dalfovo1999,leggett2001,andersen,schakel,sala-review}), 
which can be interpreted as an zero-range effective 
field theory (EFT) of the dilute Bose gas \cite{andersen,braaten,roth}. 
In the last years the EFT approach to the Bose gas 
has been extended including finite-range effects due to 
the effective range $r_s$ of the interaction potential. 
It has been shown that this approach gives a modified 
Gross-Pitaevskii equation 
\cite{sala-old,gao,boris,pethick,zinner,ketterle,sgarlata} for the condensate 
and non universal effects for quantum fluctuations 
at zero temperature \cite{andersen,braaten,roth,gao}. 

In this paper, by using finite-temperature functional 
integration \cite{schakel,atland}, we show that with a zero-range 
repulsive interaction the uniform configuration of the Bose gas 
becomes thermodynamically unstable at large values of the gas parameter 
due to Gaussian quantum fluctuations. 
This puzzling problem is solved by including 
the effects of the effective range $r_s$ of the interaction potential. 
In particular, we derive the beyond-mean-field 
(one-loop, Gaussian) equation of state of the bosonic system at 
zero temperature. This equation of state 
(see also \cite{andersen,braaten}) depends on both the scattering length $a_s$ 
and the effective range $r_s$ of the interaction. We prove that 
it is necessary a positive effective range $r_s$ above a 
critical threshold ($r_s/a_s>0.25$) to remove the thermodynamical 
instability of the uniform configuration for large values of 
the gas parameter. In the case of a hard-core interaction potential, 
where $r_s/a_s=2/3$, our EFT results for the ground-state energy 
are in quite good agreement with zero-temperature 
Monte Carlo calculations \cite{pigs}. Finally, we include thermal 
fluctuations and obtain analytically a finite-temperature equation 
of state which is reliable at low temperatures. 

\section{Partition function of the system}

In the study of the interacting Bose gas we adopt the path integral formalism, 
where bosonic atoms are described by a complex field $\psi(\mathbf{r},\tau)$. 
Within this framework, all the relevant thermodynamical properties of the 
system can be computed starting from the grand canonical 
partition function ${\cal Z}$ at finite temperature \cite{atland}
\begin{equation}
\mathcal{Z} = \int \mathcal{D}[\psi,\psi^*] \exp\bigg\lbrace 
-\frac{S[\psi,\psi^*]}{\hbar}\bigg\rbrace \;, 
\label{papo}
\end{equation}
where 
\begin{equation}
S[\psi,\psi^*] = \int_0^{\hbar \beta}d\tau\int d^3\mathbf{r} \ 
\mathcal{L}(\psi,\psi^*)
\label{action}
\end{equation}
is the Euclidean action and $\beta \equiv 1/(k_B T)$ with $k_B$ being the 
Boltzmann's constant. The grand potential $\Omega$, as a function of the 
chemical potential $\mu$ and the temperature $T$, can be obtained 
by \cite{atland}
\beq
\Omega = -\frac{1}{\beta} \log\mathcal{Z}\;.
\label{granpot def}
\eeq
By working in the grand canonical ensemble the nonlocal 
Lagrangian density of interacting identical bosons is given by 
\beq
\begin{aligned}
\mathcal{L} &= \psi^*(\mathbf{r},\tau)\bigg[\hbar\frac{\partial}
{\partial \tau} - \frac{\hbar^2\nabla^2}{2m} - \mu \bigg]
\psi(\mathbf{r},\tau) \\
& \quad +\frac{1}{2}\int d^3\mathbf{r'}|\psi(\mathbf{r'},\tau)|^2 
V(|\mathbf{r} - \mathbf{r'}|) |\psi(\mathbf{r},\tau)|^2 \;,
\end{aligned}
\label{lagrangian-initial}
\eeq
where $V(|\mathbf{r} - \mathbf{r'}|)$ is the spherically-symmetric 
two-body interaction potential between bosons. 

\section{Zero-range effective potential}

In order to get analytical results, 
several authors showed that it is worthwhile to replace the inter-atomic 
potential with a pseudo-potential which must reproduce low-energy 
scattering properties and energy shifts of the original one \cite{roth}. 
Dealing with ultracold and dilute atoms, the usual, and most simple, 
scheme consists in replacing $V(r)$ with the zero-range 
Fermi pseudo-potential $V_{p,0}(r) = g_0 \, \delta^{(3)}(r)$, 
where $\delta^{(3)}(r)$ is the Dirac delta function. Clearly, 
the Fourier transform ${\tilde V}_{p,0}(q)$ of $V_{p,0}(r)$ reads 
\beq 
{\tilde V}_{p,0}(q) = g_0 \; , 
\label{zero-range}
\eeq
where $g_0$ is given, from scattering theory, 
by $g_0 = {4\pi\hbar^2}a_s/m$ with $a_s$ 
the s-wave scattering length \cite{andersen,dalfovo1999}. 
Consequently, the Lagrangian density becomes 
\beq 
\mathcal{L} = \psi^*(\mathbf{r},\tau)\bigg[\hbar\frac{\partial}
{\partial \tau} - \frac{\hbar^2\nabla^2}{2m} - \mu \bigg]
\psi(\mathbf{r},\tau) + \frac{1}{2}g_0  |\psi(\mathbf{r},\tau)|^4  \; ,
\label{lagrangian-zero-range}
\eeq

The mean-field (saddle-point) plus Gaussian (one-loop) approximation 
is obtained setting 
\beq 
\psi({\bf r},\tau) = \psi_0 + \eta({\bf r},\tau) 
\label{shift}
\eeq
and expanding the action $S[\psi, \psi^*]$ 
of Eq. (\ref{action}) around the uniform and constant $\psi_0$ 
up to  quadratic (Gaussian) order in $\eta({\bf r},\tau)$ 
and $\eta^*({\bf r},\tau)$. 
We find that, in the momentum space, the Gaussian contribution 
of quantum fluctuation is described by
\begin{equation}
S_{g}[{\tilde \eta},{\tilde \eta}^*] = {1\over 2} \sum_{Q}
({\tilde \eta}^*(Q),{\tilde \eta}(-Q)) \ {\bf M}(Q) \left(
\begin{array}{c}
{\tilde \eta}(Q) \\
{\tilde \eta}^*(-Q)
\end{array}
\right) \;
\label{S_gaussian}
\end{equation}
where $Q=({\bf q},i\omega_n)$ is 
the $3+1$ vector denoting the momenta ${\bf q}$ and bosonic Matsubara
frequencies $\omega_n=2\pi n/(\beta \hbar)$. The matrix ${\bf M}(Q)$
is the inverse fluctuation propagator, given by the following 
\begin{widetext}
\beq
{\bf M}(Q) = \beta \,
\left(
\begin{matrix}
-i \hbar \omega_n + {\hbar^2q^2\over 2m} - \mu 
+ 2g_0\psi_0^2  &  g_0\psi_0^2 \, 
 \\
g_0 \psi_0^2 & i \hbar \omega_n 
+ {\hbar^2q^2\over 2m} -\mu + 
2g_0\psi_0^2
\end{matrix}
\right)\;.
\eeq
\end{widetext}
Integrating over the bosonic fields ${\tilde \eta}(Q)$
and ${\tilde \eta}^*(Q)$ we obtain 
the Gaussian grand potential
\beqa
\Omega_g &=& {1\over 2\beta} \sum_{Q} \ln{\mbox{Det}({\bf M}(Q))}
\nonumber
\\
&=& {1\over 2\beta} \sum_{\bf q} \sum_{n=-\infty}^{+\infty}
\ln{[\beta^2 (\hbar^2\omega_n^2+E_{\bf q}^2)]} \; ,
\label{pl}
\eeqa
where $E_{\bf q}$ is the dispersion relation:
\beq 
E_{\bf q}(\mu,\psi_0) = \sqrt{\left( {\hbar^2q^2 \over 2m} -\mu + 
2 g_0 \psi_0^2 \right)^2 - g_0^2 \psi_0^4 } \; .
\eeq
The sum over bosonic Matsubara frequencies gives \cite{andersen}
\beq
{1\over 2\beta}
\sum_{n=-\infty}^{+\infty} \ln{[\beta^2(\hbar^2\omega_n^2+E_{\bf q}^2)]} =
{E_{\bf q}\over 2} + {1\over \beta } \ln{(1-e^{-\beta E_{\bf q}})} \; ,  
\label{flavione}
\eeq
and, in this way, taking into account Eq.(\ref{lagrangian-zero-range})   
one finds the grand potential \cite{andersen,schakel,sala-review} 
\beq 
\Omega(\mu,\psi_0) = \Omega_0(\mu,\psi_0) + \Omega_g(\mu,\psi_0) + 
\Omega_{g}^{(T)}(\mu,\psi_0) \; , 
\label{maipiu}
\eeq
where 
\beq 
\Omega_0(\mu,\psi_0) = 
\left( - \mu \, \psi_0^2 + {1\over 2} g_0 \, \psi_0^4 \right) \ L^3 \;  
\label{maipiu1}
\eeq
is the mean-field contribution (assuming a real $\psi_0$) with $L^3$ the 
volume of the system, 
\beq
\Omega_{g}^{(0)}(\mu,\psi_0) = {1\over 2} \sum_{{\bf q}} E_{\bf q}(\mu,\psi_0) 
\label{maipiu2}
\eeq
is the zero-point energy of bosonic excitations, 
i.e. the zero-temperature contribution of quantum Gaussian 
fluctuations, while 
\beq
\Omega_{g}^{(T)}(\mu,\psi_0) =
{1\over \beta} \sum_{{\bf q}} \ln{\left(1 - e^{-\beta E_{\bf q}(\mu,\psi_0)} 
\right)} \; 
\label{maipiu3}
\eeq
takes into account thermal Gaussian fluctuations.

Imposing the crucial condition 
\beq 
{\partial \Omega_0(\mu,\psi_0)\over \partial \psi_0} = 0 \; , 
\label{chesara}
\eeq
one gets 
\beq 
\psi_0(\mu) = \sqrt{\mu\over g_0} 
\label{mu0}
\eeq 
and the well-known Bogoliubov spectrum of collective excitations  
\beq 
E_{\bf q}(\mu) = \sqrt{ {\hbar^2q^2 \over 2m} 
\left( {\hbar^2q^2 \over 2m} + 2\mu \right) } \; .
\eeq
Remarkably, the spectrum is now gapless as required by the Goldstone 
theorem. Now, one can replace $\psi_0$ with its classical value given 
by Eq. \eqref{mu0} and, after removing the ultraviolet divergence 
(for a detailed review see \cite{sala-review}) in 
$\Omega_g^{(0)}(\mu)=\Omega_g^{(0)}(\mu,\psi_0(\mu))$, 
the zero temperature grand potential becomes  
\beq 
\Omega_{g}^{(0)}(\mu) = 
{8\over 15 \pi^2} \bigg({m\over \hbar^2}\bigg)^{3/2} \mu^{5/2} \; . 
\label{omega_g_zero}
\eeq
The pressure $P(\mu)$ of the system is simply related to the 
grand potential $\Omega(\mu)$ by the formula 
\beq
P(\mu) = -{\Omega(\mu) \over L^3}  \; .  
\eeq
At zero temperature the beyond-mean-field pressure $P(\mu)$ as a function 
of the chemical potential $\mu$ is then given by 
\beq 
P(\mu) = {\mu^2\over 2\, g_0} 
- {8\over 15 \pi^2} \bigg({m\over \hbar^2}\bigg)^{3/2} \mu^{5/2} \; ,    
\label{p-unstab}
\eeq 
and it agrees with the result derived in 1960 by Lee and Yang \cite{lee1960}  
within the framework of quantum statistical mechanics. 
Eq. (\ref{p-unstab}) has been also obtained in \cite{haugset} by using 
of a functional approach. It is important to stress that the 
pressure $P$ given by Eq. (\ref{p-unstab}) becomes negative, 
i.e. unphysical, for a large value of the chemical potential $\mu$. 
Actually, the uniform configuration is thermodynamically 
stable if and only if 
\beq
\frac{\partial^2 P(\mu)}{\partial \mu^2} > 0  \; .
\label{thermostab}
\eeq 
Eq. (\ref{p-unstab}) implies thermodynamical 
instability for $\mu > \mu_c = \pi (\hbar^2/(2m))^{3/4}/\sqrt{2g_0}$. 

The zero-temperature number density $n$ can be derived from 
the grand potential by using this thermodynamic formula 
\beq 
n(\mu) = - {\partial \Omega(\mu) \over \partial \mu} \; , 
\label{number-equation}
\eeq
which gives 
\beq
n(\mu) = {\mu \over g_0} - {4\over 3\pi^2} 
\left({m\over \hbar^2}\right)^{3/2} \mu^{3/2} \; . 
\label{camilla}
\eeq
From this equation one can easily numerically 
determine $\mu$ as a function of $n a_s^3$. The result is 
the solid line of Fig. 1. The plot clearly shows that 
in the absence of finite-range corrections the chemical 
potential has two branches and there are no solutions above 
a critical value of the gas parameter $n a_s^3$.  
This result, that is full consistent with the discussion 
of Eq. (\ref{thermostab}), means that for $n a_s^3> 0.004$ the uniform 
configuration does not exist anymore. 

We observe that a simple analytical result is 
obtained from Eq. (\ref{camilla}) by using a perturbative expansion 
where $g_0 n \ll \mu$. In this way one gets
\beqa 
\mu(n) &=& g_0 n + {4g_0 \over 3\pi^2} 
\left({m\over \hbar^2}\right)^{3/2} (g_0 n)^{3/2} \; . 
\label{camilla1}
\eeqa
The first term of this chemical potential was derived by Bogoliubov 
\cite{bogoliubov} while the second term is the one deduced by Lee, 
Huang and Yang \cite{yang}. However, it is extremely important to remind 
that Eq. (\ref{camilla1}) is obtained assuming a very small 
gas parameter $n a_s^3$ (parturbative scheme above mean-field 
plus Gaussian results) and it cannot be used when the system is 
thermodynamically unstable according to the general 
equation Eq. (\ref{camilla}). Close to the instability 
the contribution of Gaussian fluctuations to the grand potential, 
given by Eq. \eqref{omega_g_zero}, becomes of the same order 
of the mean-field term; this signals that quantum fluctuations are 
strong enough to destabilize the uniform configuration. 
The critical value  $(na_s^3)_c \simeq 0.004$ specifies the upper 
threshold of applicability of the zero-range 
Gaussian theory and, consequently, of Eq. \eqref{camilla1}.

\begin{figure}[t]
\centerline{\epsfig{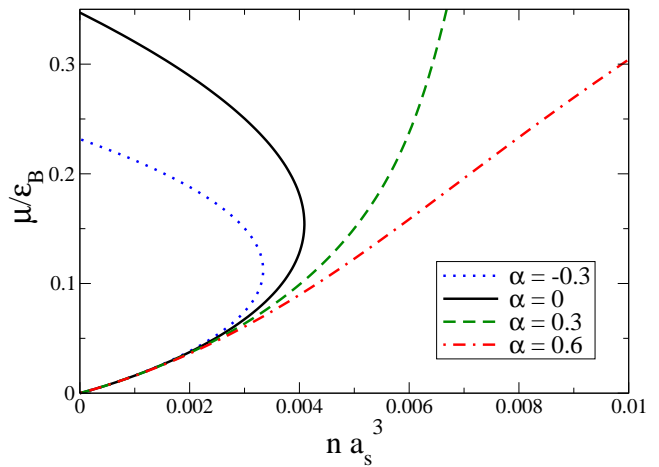}}
\small 
\caption{(Color online). Chemical potential $\mu$ vs gas parameter $n a_s^3$ 
obtained with beyond-mean-field (Gaussian) EFT, Eq. (\ref{tegico}), 
for different values of the adimensional ratio $\alpha=2 r_s/a_s$. 
Here $\epsilon_B=\hbar^2/(m a_s^2)$ is the characteristic 
energy of the interacting Bose gas, $a_s$ is the s-wave scattering 
length, $r_s$ is the effective range, and $n$ is the number density.} 
\label{fig1}
\end{figure} 

We shall now show that the inclusion of a effective-range interaction 
in the thermodynamics can remove the instability 
by assuring a positive sign of the second derivative of the pressure. 

\section{Finite-range effective potential}

An improvement of the contact (zero-range) approximation can be achieved 
by replacing the interaction potential ${\tilde V}({\bf q})$ 
with the finite-range pseudo-potential 
\begin{equation}
\tilde{V}_{p,2}(q) = g_0 + g_2 \ q^2 \; .
\label{pseudo-potential q}
\end{equation}
The relation with the true inter-atomic potential appearing in 
Eq. \eqref{lagrangian-initial} is given by the following relations 
\begin{equation}
g_0 = \tilde{V}(0) =  \int d^3\mathbf{r}\ V(r) 
\label{g0_VR}
\end{equation}
and
\begin{equation}
g_2 = \frac{1}{2}\tilde{V}''(0) = 
-\frac{1}{6}\int d^3\mathbf{r} \ r^2 \ V(r) 
\label{g2_VR}
\end{equation}
where $\tilde{V}(\mathbf{q}) = 
\int d^3 \mathbf{r} \exp(i\mathbf{q}\cdot\mathbf{r})V(r)$. 
It has been shown \cite{gao,roth} that, in real space, 
the pseudo-potential in \eqref{pseudo-potential q} is given by 
\begin{equation}
V_{p}(r) = g_0 \, \delta(r) - {g_2\over 2} \bigg[ {\overleftarrow\nabla}^2 
\, \delta(r) + \delta(r) \, {\overrightarrow\nabla}^2 \bigg] \; .
\label{pseudo-potential r}
\end{equation}
 
The connection with experimental quantities such as the s-wave 
scattering length $a_s$ and the s-wave effective range $r_s$ 
can be established by requiring the matching between the expansion 
parameters of Eq. (\ref{pseudo-potential q}) and the ones obtained 
by a more general pseudo-potential $V_p(q)$. 
Several authors \cite{andersen,braaten,roth} adopt 
\begin{equation}
\tilde{V}_p(q) = \frac{4\pi\hbar^2}{m}\frac{\tan{(\delta_0(q)})}{q} \; , 
\label{pseudo-potential-roth}
\end{equation}
which depends on the s-wave phase shift $\delta_0(q)$. 
Notice that $\delta_0(q)$ is related to the scattering length 
$a_s$ and the effective range $r_s$ by the equation
\begin{equation}
\delta_0(q) = \arctan\bigg(\frac{1}{-\frac{1}{a_s} + 
\frac{1}{2}r_s q^2 + O(q^4)}\bigg) \;.
\label{phase shift}
\end{equation}
One can expand $\tilde{V}_p(q)$ and $\delta_0(q)$, respectively Eq. 
\eqref{pseudo-potential-roth} and Eq. \eqref{phase shift}, up to the second 
order for small $q$. In this way, one finds that 
the coupling constants $g_0$ and $g_2$ are related to the physical 
parameters $a_s$ and $r_s$ according to
\begin{equation}
g_0 = \frac{4\pi\hbar^2}{m}a_s
\label{g_0}
\end{equation}
which is a well-known relation, and 
\begin{equation}
g_2 = \frac{2\pi\hbar^2}{m}a_s^2 r_s \; .
\label{g_2 braaten}
\end{equation}

By using the pseudo-potential in Eq. (\ref{pseudo-potential q}) 
the nonlocal Lagrangian density given by Eq. \eqref{lagrangian-initial} becomes 
\beq
\begin{aligned}
\mathcal{L} & = \psi^*(\mathbf{r},\tau) \bigg[\hbar\frac{\partial}
{\partial \tau} - \frac{\hbar^2 \nabla^2}{2m} - \mu \bigg]
\psi(\mathbf{r},\tau) \\
&\;\; + \frac{g_0}{2}|\psi(\mathbf{r},\tau)|^4 - \frac{g_2}{2}|
\psi(\mathbf{r},\tau)|^2 \nabla^2 |\psi(\mathbf{r},\tau)|^2 \;.
\end{aligned}
\label{lagrangian-3}
\eeq
This is the finite-range effective field theory (EFT) we shall use 
in the remaining part of the paper. Gaussian (one-loop) 
results of Eq. (\ref{lagrangian-3}) have been obtained 
in Refs. \cite{braaten,andersen,roth}, but mainly in the 
perturbative regime within one-loop calculations 
and at zero temperature. Here we explicitly prove that, 
working in regions where the gas parameter is small, 
the inclusion of the effective-range term can remove the 
instability of the zero-range theory. Moreover, we compare our 
EFT calculatons with Monte Carlo data and analyze also finite-temperature 
effects. 

By using the Lagrangian density derived in Eq. \eqref{lagrangian-3} 
in the Eq. \eqref{action}, 
the stationary Gross-Pitaevskii equation for the space-dependent 
field $\psi_0(\mathbf{r})$ 
can be derived by means of the saddle-point approximation
\begin{equation}
\delta S\big[\psi^*_0(\mathbf{r}),\psi_0(\mathbf{r})\big] = 0 \; ,
\label{stazionarizzazione}
\end{equation}
which leads to 
\begin{equation}
\bigg[-\frac{\hbar^2 \nabla^2}{2m} + g_0\big| \psi_0(\mathbf{r})\big| 
-g_2\nabla^2\big|\psi_0(\mathbf{r})\big|^2 \bigg]\psi_0(\mathbf{r}) = 
\mu \psi_0(\mathbf{r}) \;.
\label{mgpe}
\end{equation}
Eq. \eqref{mgpe} was derived for the first time in \cite{gao} 
for a Bose gas under external confinement. 
In this case, simulations based on Eq.\eqref{mgpe} leads to a better 
agreement with Quantum Monte Carlo datas concerning, for example, 
the grounde-state energy \cite{gao}. Moreover, by using Eq.\eqref{mgpe} 
it is possible to study, also in the absence of an external confining 
potential, how space dependent topological solutions, 
such as vortex and solitons, are affected by this 
effective-range expansion \cite{sgarlata}.

As in the previous section, in the remaining part of the paper 
we adopt the shift of Eq. (\ref{shift}) 
and expand the action $S[\psi, \psi^*]$ of Eq. (\ref{action}) around a 
uniform and stationary $\psi_0$ 
up to  quadratic (Gaussian) order in $\eta({\bf r},\tau)$ 
and $\eta^*({\bf r},\tau)$, but now using 
(\ref{lagrangian-3}) instead of (\ref{lagrangian-zero-range}). 
Formally, we find again Eqs. (\ref{maipiu}), (\ref{maipiu1}), 
(\ref{maipiu2}), (\ref{maipiu3}) for the grand potential $\Omega$ 
but now the dispersion relation reads 
\beqa
E_{\bf q}(\mu,\psi_0) &=&
\Big[ \bigg( {\hbar^2q^2 \over 2m} -\mu
+ \psi_0^2 (g_0+{\tilde V}_{p,2}(q)) \bigg)^2 
\nonumber 
\\
&-& \psi_0^4 {\tilde V}_{p,2}(q)^2 \Big]^{1/2} \; .
\label{spettrob0}
\eeqa

By using Eqs. (\ref{chesara}) and (\ref{mu0}) to remove 
the dependence on $\psi_0$ in $E_{\bf q}$ we obtain 
\beq 
E_{q}(\mu) = \sqrt{{\hbar^2q^2 \over 2m}
\left( (1+\chi \mu) 
{\hbar^2q^2 \over 2m} + 2 \mu \right)} \; ,  
\label{bogo2}
\eeq
where 
\beq 
\chi = {4m\over \hbar^2} {g_2\over g_0} \;  
\label{chi}
\eeq
takes into account finite range effects of the inter-atomic potential. 

\subsection{Zero-temperature results}

The zero-temperature Gaussian grand potential 
$\Omega_{g}^{(0)}(\mu)$, given by Eq. (\ref{maipiu2}),  
is ultraviolet divergent with $E_{q}(\mu)$ given by Eq. (\ref{bogo2}). 
However, this divergence can be regularized with 
dimensional regularization \cite{thooft}. For a recent review of this and other 
regularization methods applied to the dilute and ultracold atomic systems 
one can see \cite{sala-review}. In this way we find 
\beq 
{\Omega_{g}^{(0)} \over L^3} = {8\over 15\pi^2} 
\left({m\over \hbar^2}\right)^{3/2} {\mu^{5/2} \over 
(1+\chi \mu)^2}  \; .   
\label{omegag0}
\eeq 

The zero-temperature number density $n$ is obtained by using the 
number equation (\ref{number-equation}) with $\Omega$ given by Eq. 
(\ref{maipiu}) with Eqs. (\ref{maipiu1}), (\ref{maipiu2}) 
and (\ref{omegag0}). We obtain 
\beqa 
n(\mu) &=& {\mu \over g_0} - {4\over 3\pi^2} 
\left({m\over \hbar^2}\right)^{3/2} {\mu^{3/2} \over 
(1+\chi \mu)^2} 
\nonumber 
\\
&+& {64\over 15\pi^2} 
\left({m\over \hbar^2}\right)^{5/2} {g_2\over g_0} 
{\mu^{5/2} \over (1+\chi \mu)^3} \; . 
\label{tegico}
\eeqa
From this equation we determine $\mu$ as a function of $n a_s^3$ for different 
values of $\chi$ at fixed density $n$. The results are shown in Fig. 1, 
where $\alpha$ is proportional to the ratio $g_2/g_0$ in adimensional 
units, namely 
\beq 
\alpha = {\hbar^2\over m a_s^2} \chi = 
4 {g_2\over g_0 a_s^2} = 2 {r_s\over a_s} \; . 
\label{alpha-def}
\eeq
As previously discussed, Fig. (\ref{fig1}) clearly shows that 
in the absence of finite-range corrections ($\alpha =0$) the chemical 
potential $\mu$ versus $n$ has no solutions above 
the critical value $0.004$ of the gas parameter $n a_s^3$. 
This problem is indeed solved by using 
a positive value of $\alpha$ larger than about $0.25$, while for $\alpha<0$ 
the problem gets worse. 
The different behaviour of numerical solutions of Eq. \eqref{tegico} for 
$\alpha \neq 0$ compared to Eq. \eqref{camilla} (equivalent to 
$\alpha = 0$ case) can be understood thanks to modified dependence 
from $\mu$ of Eq. \eqref{omegag0}: for $\alpha \gtrsim 0.25$ 
the finite-range Gaussian correction never becomes of the same order, 
or bigger, of the mean field term. Differently from the zero-range case, 
the finite-range correction manages to control 
the growth of fluctuations and it stabilizes the system, as highlighted 
by the dashed and dashed-dotted lines in Fig. (\ref{fig1}): for these 
values of $\alpha$ there is no critical value of the gas parameter. 

\begin{figure}[t]
\centerline{\epsfig{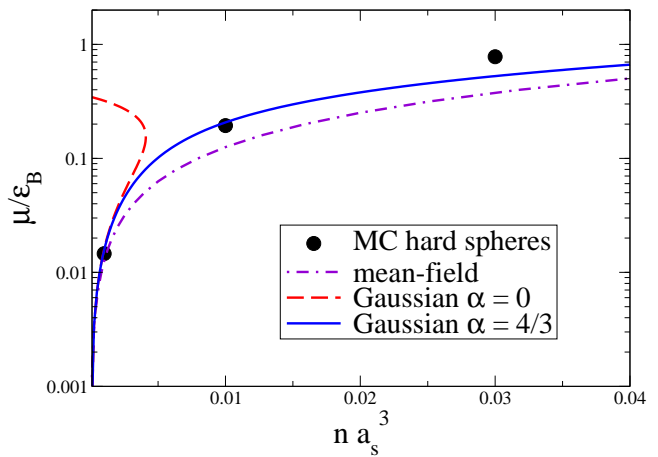}}
\small 
\caption{(Color online). Monte Carlo (MC) data (filled circles) of the 
chemical potential $\mu$ vs gas parameter $n a_s^3$ for a Bose gas of 
hard spheres \cite{pigs}. Dot-dashed line is the mean-field theory, 
$\mu= 4\pi\hbar^2a_sn/m$. Solid and dashed lines are the 
results of our zero-temperature Gaussian EFT, Eq. (\ref{tegico}), 
for two different values of the effective-range adimensional 
parameter $\alpha =2 r_s/a_s$. The case $\alpha = 0$ corresponds 
to usual scheme with a zero-range interaction while $\alpha=4/3$ 
correspond to the case of a hard-core interaction potential. 
Also here $\epsilon_B=\hbar^2/(m a_s^2)$ is the characteristic 
energy of the interacting Bose gas, $a_s$ is the s-wave scattering 
length, $r_s$ is the effective range, and $n$ is the number density.} 
\label{fig2}
\end{figure}

It is interesting to compare it with Quantum Monte Carlo (QMC) data 
obtained for a Bose gas of hard spheres 
by using the path-integral ground-state method \cite{pigs}, 
as done in Fig. \ref{fig2}. In the case of hard spheres 
the s-wave scattering length $a_s$ and effective 
range $r_s$ are related by $r_s/a_s=2/3$ (see, for instance, 
\cite{gao,pethick,sgarlata}). The adimensional parameter  
$\alpha$, Eq. (\ref{alpha-def}), of our theory must be $\alpha =4/3$ 
in order to model the hard-sphere Bose gas. The figure shows 
that the Monte Carlo data (filled circles) are reproduced 
reasonably well by our EFT with $\alpha=4/3$ (solid line) also 
for quite large values of the gas parameter $n a_s^3$. So, by including the 
finite-range Gaussian corrections in Eq. \eqref{tegico},
we are able to recover the zero-range results, which is reliable at very 
low values $na_s^3$ and surely not above $na_s^3 \simeq 0.004$, but we can 
reproduce the QMC datas for a more dense system. The range of applicability 
of our finite-range theory is of the order shown in the horizontal axis of 
Fig. (\ref{fig2}). 

QMC simulations are performed with a hard-core potential 
which has a finite range. Our theory, at a Gaussian level, 
introduces an effective range $r_s$ whose relation with the 
scattering length $a_s$ of the hard-core potential is $r_s = (2/3) a_s$. 
In the case of the hard-core potential, at fixed scattering 
length $a_s$ (and consequently at fixed effective range $r_s$), 
by increasing the gas parameter $n a_s^3$ one increases the density $n$ 
and, at the same time, the average distance $d \simeq n^{-1/3}$ 
between atoms reduces. In this way, $d$ becomes comparable 
with the effective range $r_s$ and finite-range effects of the 
inter-atomic potential are then sizable. 

We stress that, at two-loop level (next-to-Gaussian), the correction to 
energy density proportional to $na_s^3$ is characterized by an unknown 
coefficient which cannot be specified in the zero-range framework \cite{nieto}. 
This coefficient can be expressed in terms of the s-wave scattering 
length and a three-body coupling not easy to determine experimentally. 
It is remarkable that, at least for alkali atoms, second-order 
quantum corrections are proportional to the logarithm of an additional  
length scale fixed by the van der Waals interaction.

\subsection{Finite-temperature results}

The finite-temperature one-loop contribution to the equation 
of state is obtained from $\Omega_g^{(T)}(\mu)$, which can be written as 
\beq 
{\Omega_g^{(T)}(\mu)\over L^3} = 
-{1\over 6\pi} \int_0^{\infty} dq \, q^3 \, {dE_q\over dq} 
{1\over e^{\beta E_q(\mu)}-1} \; . 
\eeq
Introducing the variable $x=\beta E_q(\mu)$ we get 
\beq 
{\Omega_g^{(T)}(\mu)\over L^3} = -{1\over 6\pi^2\beta} \int_0^{\infty} dx \, 
q(x,\mu)^3 {1\over e^x -1} \; , 
\eeq
where $q(x,\mu)$ is given by 
\beq 
q(x,\mu) = \sqrt{2m \mu \over \hbar^2 (1+\chi \mu)} \sqrt{-1 + \sqrt{1+ 
{(1+\chi \mu) x^2\over \mu^2\beta^2}}} \; . 
\eeq
Expanding this expression at low temperature $T$ we find 
\beq 
{\Omega_g^{(T)}(\mu) \over L^3} = - {\pi^2\over 90} ({m\over \hbar^2})^{3/2} 
{(k_BT)^4\over \mu^{3/2}} \left( 1 - {5\pi^2 \over 7} (k_B T)^2 
{1+\chi \mu \over \mu^2} \right) \; ,  
\label{ordoab}
\eeq
and the finite-temperature contribution $n_g^{(g)}$ to the total number  
density $n$ reads 
\beq 
n_g^{(T)}(\mu) = - {\pi^2\over 60} ({m\over \hbar^2})^{3/2} 
{(k_BT)^4\over \mu^{5/2}} \left( 1 - {5 \pi^2\over 21} (k_BT)^2 
{7 + 5\chi \mu \over \mu^2} \right) \; . 
\eeq
Thus, within the perturbative expansion approach 
($|\mu - g_ n| \ll 1$) previously discussed, 
the finite-temperature equation of state reads 
\beq 
\mu(n) = \mu_0(n) + \mu_g^{(0)}(n) + \mu_g^{(T)}(n) \; , 
\label{chaos}
\eeq
where $\mu_0(n) + \mu_g^{(0)}(n)$ is given by Eq. (\ref{camilla1}) and   
\beqa
\mu_g^{(T)}(n) &=& - {\pi^2\over 60} g_0 ({m\over \hbar^2})^{3/2} 
{(k_BT)^4\over (g_0 n)^{5/2}} \Big[ 1 - {5\pi^2\over 21} (k_BT)^2 
\nonumber 
\\
&\times&  {7 + 5\chi (g_0n) \over (g_0n)^2} \Big] \; . 
\label{camilla3}
\eeqa
Notice that Eq. (\ref{chaos}), with Eqs. (\ref{camilla1}) and (\ref{camilla3}), 
generalizes the old familiar result obtained in 1958 by 
Lee and Yang \cite{lee}.

\section{Conclusions}

We have used a finite-temperature Gaussian (one-loop) functional integration 
to obtain the equation of state for a dilute and ultracold gas 
of bosons with uniform number density $n$, 
taking into account both the scattering length $a_s$ and 
the effective range $r_s$ of the inter-atomic interaction. 
The divergent zero-point energy 
of the system has been regularized by performing dimensional 
regularization. Our analytical results at zero and finite temperature, 
which are non trivial generalizations of old but familiar formulas 
\cite{bogoliubov,lee1960,yang,lee} depending only 
on the scattering length $a_s$, are in quite good agreement with 
recent Monte Carlo calculations \cite{pigs} also for relatively 
large values of the gas parameter $na_s^3$. 
As discussed by Braaten, Hammer and Hermans \cite{braaten}, 
the Gaussian grand potential can be improved 
taking into account also two-loop corrections but, in this case, 
a three-body interaction is needed to regularize 
the divergent two-loop grand potential \cite{braaten,nieto}. 
On this respect a self-consistent derivation of an effective field theory 
with a three-body term starting from a nonlocal 
two-body interaction potential is still missing and it surely 
deserves a deep investigation. 

\section*{Acknowledgments}

The authors acknowledge for partial support Ministero Istruzione 
Universita Ricerca (PRIN Project "Collective Quantum Phenomena: from 
Strongly-Correlated Systems to Quantum Simulators"). The authors 
thank Prof. Flavio Toigo for many enlightening discussions.

\end{document}